\documentclass{article}

\usepackage{arxiv}

\usepackage[utf8]{inputenc} % allow utf-8 input
\usepackage[T1]{fontenc}    % use 8-bit T1 fonts
\usepackage{hyperref}       % hyperlinks
\usepackage{url}            % simple URL typesetting
\usepackage{booktabs}       % professional-quality tables
\usepackage{amsfonts}       % blackboard math symbols
\usepackage{nicefrac}       % compact symbols for 1/2, etc.
\usepackage{microtype}      % microtypography
\usepackage{lipsum}		% Can be removed after putting your text content
\usepackage{graphicx}
\usepackage{natbib}
\usepackage{doi}

\title{Trajectories of long duration balloons launched 
from McMurdo Station in Antarctica}

\author{ \hspace{1mm} Christopher Geach \\
	German Aerospace Center\\
	Neuestrelitz, Germany \\
	\texttt{christopher.geach@dlr.de} \\
	\And
	Shaul Hanany\\
	School of Physics and Astronomy\\
	University of Minnesota, Twin-Cities\\
	Minneapolis, MN, USA \\
	
	\And
	Chiou Yang Tan\\
	School of Physics and Astronomy\\
	University of Minnesota, Twin-Cities\\
	Minneapolis, MN, USA \\
	
	\And
	Xin Zhi Tan\\
	Department of Physics\\
	University of Illinois Urbana-Champaign\\
	Urbana, IL, USA \\
}

\hypersetup{
pdftitle={Trajectories of long duration balloons launched 
from McMurdo Station in Antarctica},
pdfsubject={},
pdfauthor={Christopher Geach, Shaul Hanany, Chiou Yang Tan, Xin Zhi Tan},
pdfkeywords={Stratospheric ballooning, long-duration ballooning, Antarctica, McMurdo, trajectories},
}

\begin{document}
\maketitle

\begin{abstract}
The Columbia Scientific Ballooning Facility operates stratospheric balloon flights out of McMurdo Station in Antarctica. We use balloon trajectory data from 40 flights between 1991 and 2020 to give the first quantification of trajectory statistics. We provide the probabilities as a function of time for the payload to be between given latitudes, and we quantify the southernmost and northernmost latitudes a payload is likely to attain. We find that for a flight duration of 18~days, there is 90\% probability the balloon would drift as far south as $88^{\circ}$S or as far north as $71^{\circ}$S; shorter flights are likely to experience smaller ranges in latitude. These statistics, which are available digitally in the public domain, will enable scientists planning future balloon flights make informed decisions during both mission design and execution. 
\end{abstract}

% keywords can be removed

\section{Introduction}
\label{sect:intro}  % \label{} allows reference to this section
Long duration balloon (LDB) flights have been launched by NASA
from McMurdo Station in Antarctica since the early 1990's. With the logistical support of NSF's Office of Polar Programs, more than 50 large scientific payloads have flown to date, addressing a wide array of science topics encompassing the structure and evolution of the Universe~\cite{ebex}, the nature of high energy particles~\cite{supertiger,anita}, the origins of solar activity~\cite{FGE, GRIPS}, the precipitation of magnetospheric electrons~\cite{MAXIS}, and the processes controlling the formation of stars in the Milky Way~\cite{blast}.

In the Antarctic summer, the weakened polar vortex carries stratospheric balloons westward on a generally circumpolar trajectory. For the majority of time, the balloon and payload are above the Antarctic continent. Presently, balloon flights circumnavigating Antarctica give the longest possible balloon flights on the planet with high degree of confidence of payload recovery. In 2013, the SuperTIGER payload circumnavigated the continent nearly three times to make a record flight length of 55 days. Since flights from New Zealand use only super-pressure balloons which are limited to science payload weights near 3000 lbs, Antarctic flights are the only long-flight option available for heavier payloads. 

When preparing for a LDB flight, it is often necessary to estimate the range of trajectories the payload may take. For example, thermal modeling depends on the elevation of the sun, which depends on latitude~\cite{thermal_model_paper}; for astrophysical experiments, regions of the sky available for observation depend on payload location~\cite{blast}; the trajectory of the flight will influence the strength and frequency of balloon-based observations of aurora~\cite{aurora_balloon} and of polar mesospheric clouds~\cite{pmc}. In all cases, more efficient mission planning and real time response can be made with higher confidence constraints on the trajectory of the payload. However, no quantitative information on trajectory statistics is available.

Rudimentary visual information about trajectories for 17 flights was given in 2004 by Gregory and Stepp\cite{greg_stepp}. Since then, the number of flights has tripled.  In this paper we give trajectory statistics using LDB flight data recorded over the last 30 years ending in 2020. These are the first statistics to be conducted on Antarctic balloon flight data. The motion of the balloons is primarily to the west, with relatively small deviations to the north and south. Since latitude is the driver of the majority of the trajectory-dependent effects discussed above, we concentrate on latitude data in order to quantify the range of latitudes a flight of a given length is likely to experience.

\section{Data}
\label{sec:data}

The Columbia Scientific Ballooning Facility provided trajectory data for 51 balloon flights that launched from McMurdo between 1991 and 2020. In one case, higher resolution data were provided by the ANITA science team. These data nominally include latitude, longitude, and in most cases altitude information, irregularly sampled on timescales ranging from 0.5 to 50~minutes (except for one outlier with 150~minute sampling). We excluded 10 flights from our analysis because they never reached a stable float altitude or because altitude data were not included, as well as the one flight with particularly slow sampling rate. Among the 40 remaining flights, intermittent gaps of several hours are common, particularly among earlier flights. The longest such gap is  2.7~days; gaps of 2~hours or longer comprise 5.3\% of the total time these flights were
airborne.

\begin{figure}
\centering
\includegraphics[width=16cm]{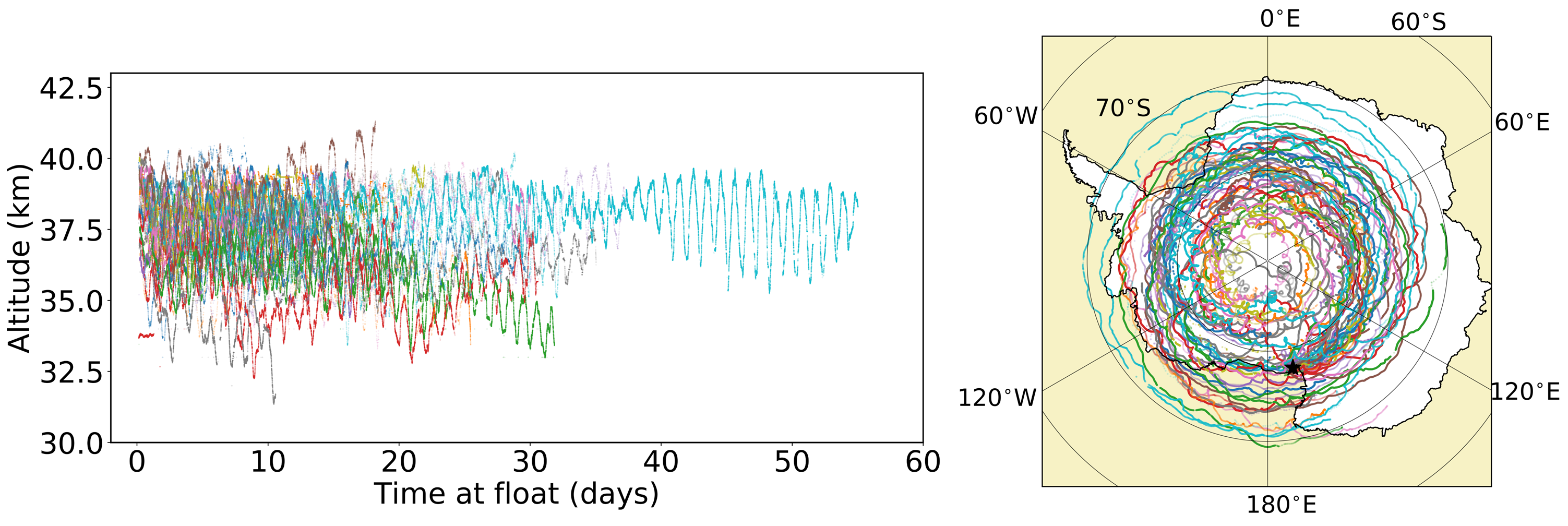}
\caption{Altitude as a function of time for the 40 LDB flights that are used to provide trajectory statistics (left panel) and a map of their trajectories (right panel). A black star indicates the location of McMurdo Station.  
\label{fig:figure1} }
\end{figure}

\begin{figure}
\begin{center}
\includegraphics*[width=8cm]{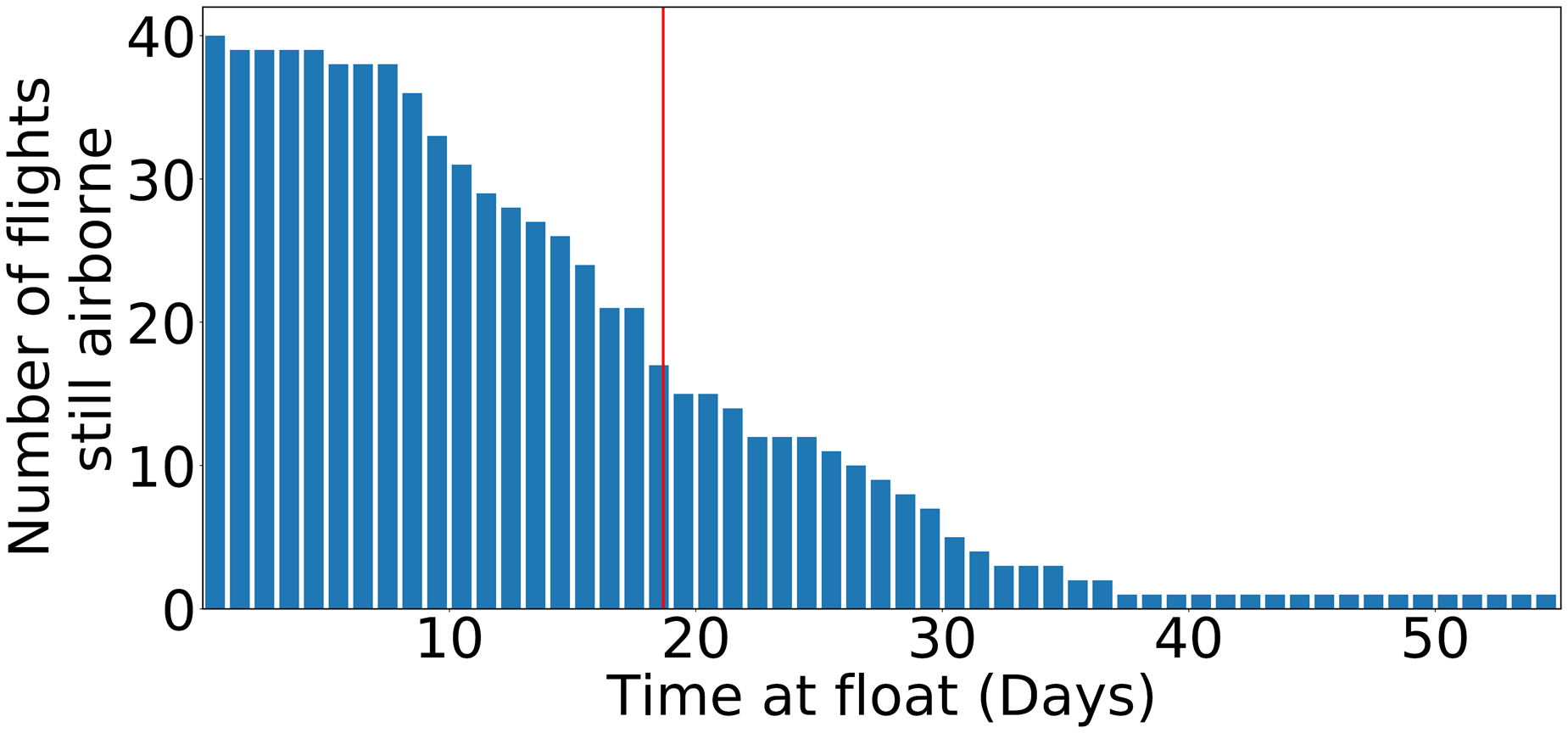}
\end{center}
\caption{Number of flights remaining airborne at least up to the given time at float. The median flight duration was 18.7 days (red vertical line).
\label{fig:figure2}}
\end{figure}

The left panel of Figure~\ref{fig:figure1} gives the altitudes of the 40 remaining flights as a function of time. The mean of the altitude data is 37.3~km, and 90\% of the data is between 34.7 and 39.3~km. 
Thus, the data are dominated by wind patterns within a $\sim$4.5~km layer of the atmosphere. The minimum and maximum altitudes are 31.4 and 41.3~km, respectively. The right panel of Figure~\ref{fig:figure1} and Figure~\ref{fig:figure2} show the trajectories and the distribution of durations of these 40 flights, respectively. The median duration was 18.7~days. The balloons had median drift speeds between 4 and 12~m/s and drifted west at an average rate of 15 to 45 degrees longitude per day.

\section{Analysis}

We analyzed the data to provide three products: (1) the probability that a particular pixel in latitude/longitude was visited, (2) the probability as a function of time to be in a particular latitude bin, and (3) the probability to attain a given minimum and maximum latitude throughout flight. To generate these products, we performed a linear interpolation in longitude and latitude over gaps in the data; we discuss the errors associated with this interpolation at the end of this section.

\subsection{Probability of pixel visitation}

\begin{figure}
\begin{center}
\includegraphics*[width=10cm]{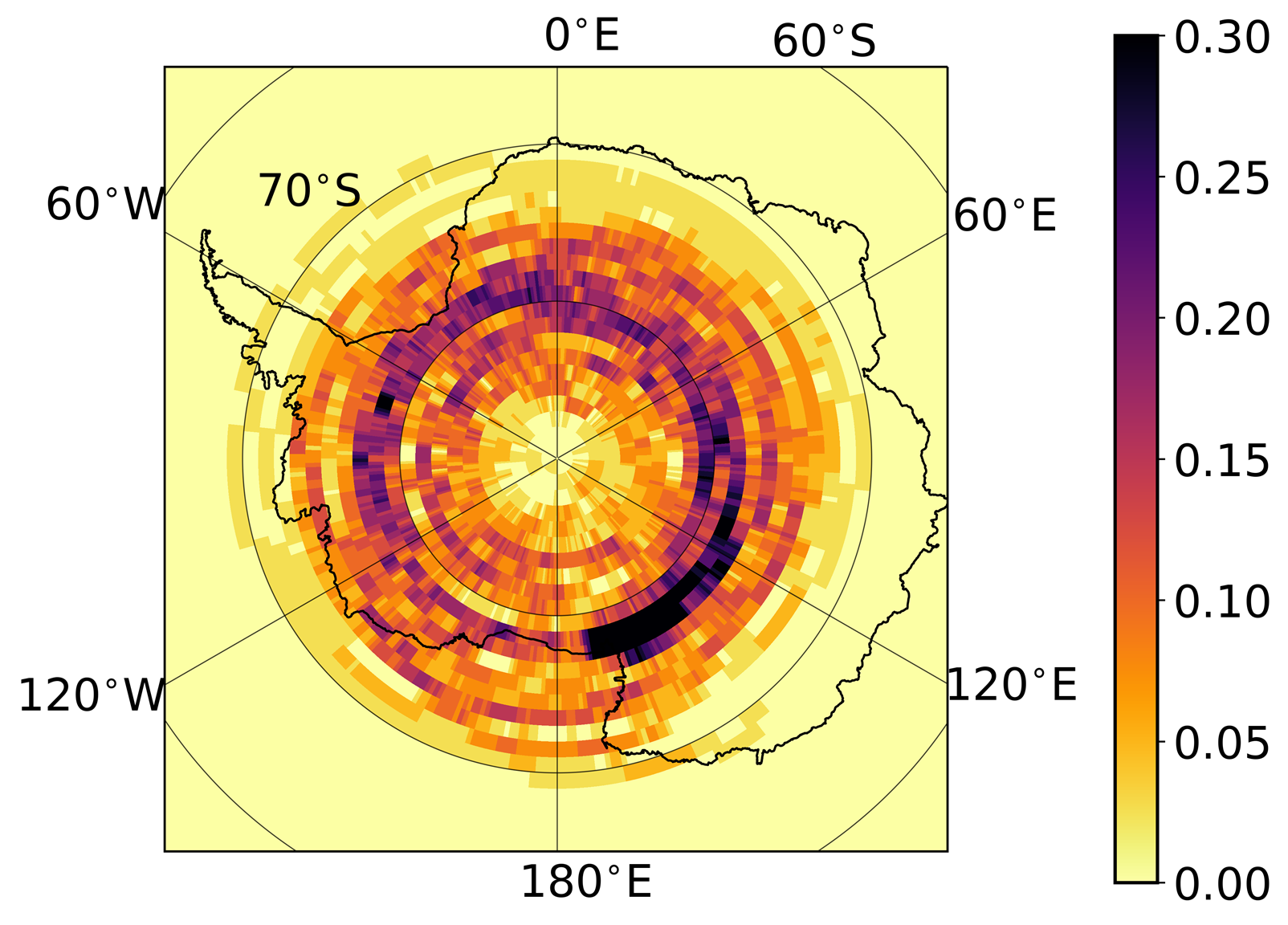}
\end{center}
\caption{Hitmap of balloon trajectories. Each pixel is colored according to the fraction of flights that passed through it. The color scale saturates at 0.30 to improve contrast; many bins around the launch location are over-saturated.
\label{fig:figure3} }
\end{figure}

We created a grid in latitude and longitude with $1^{\circ}$ bins in each direction. We
counted the number of flights that passed through each bin, then normalized by the total
number of flights to derive an inferred probability of a flight visiting the given pixel. 
This probability map is plotted in Figure~\ref{fig:figure3}, with color indicating the probability of a flight passing through a given pixel. Note that these are not equal-area pixels, as they become progressively narrower near the pole. We ignore this effect, as the predominant motion of the balloons is to the west, and the extent of all pixels is equal in the perpendicular (north-south) direction.

\subsection{Latitude probability as a function of time} 
\label{sec:lat_vs_time}

\begin{figure} 
\begin{center}
\includegraphics*[width=10cm]{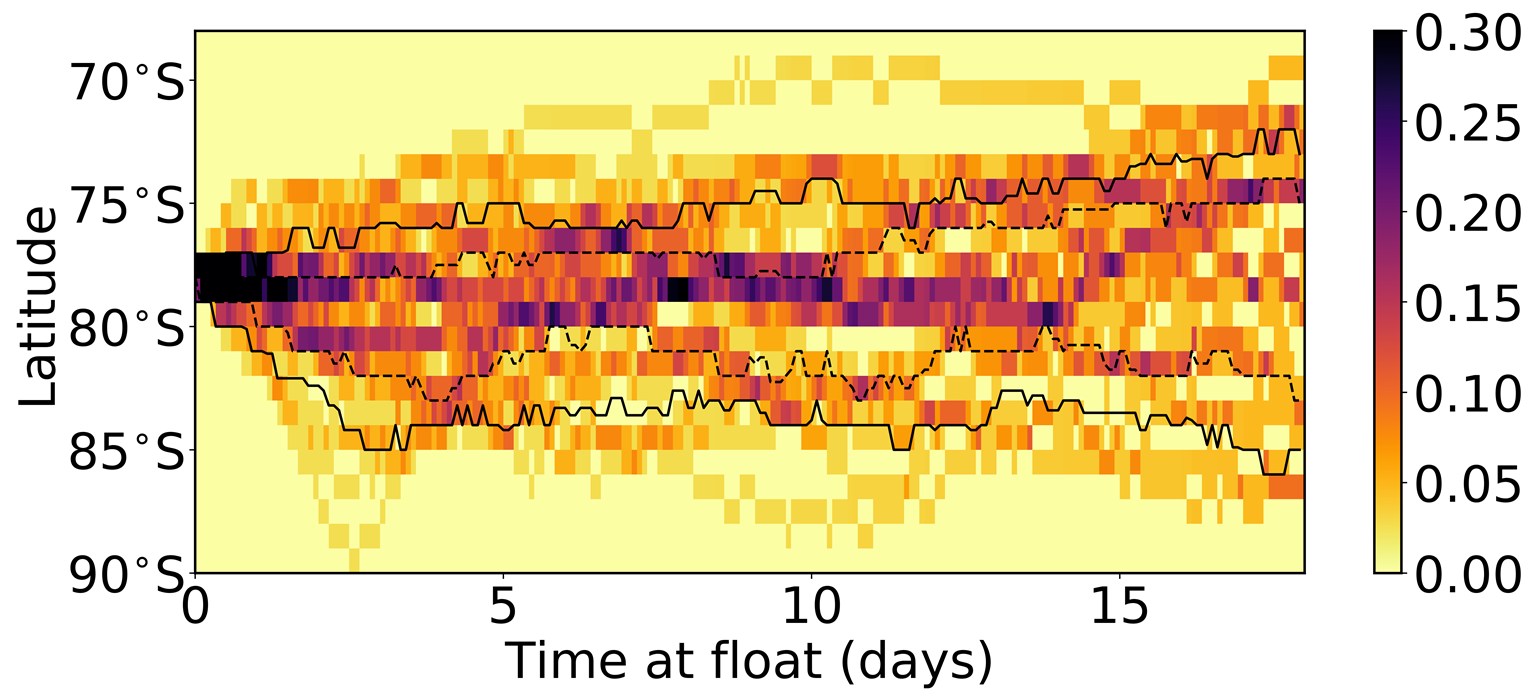}
\end{center}
\caption{Latitude probability as a function of time from launch. The solid contours give
the 90th and 10th percentile, and the dashed contours give the 75th and 25th percentile for
each temporal bin.
\label{fig:figure4}}
\end{figure}

Chronology information is missing from the relative probability displayed in
Figure~\ref{fig:figure3}. To provide time dependence information, we binned and averaged
the latitude data of each flight in 2~hour intervals starting from launch. In the vast
majority of cases, the averaging is over a range of latitudes of less than $1^{\circ}$.  In
the 6400 2-hour time segments for all 40 balloon flights, a change of $1^{\circ}$ or more
was observed only once. For each temporal bin we histogrammed the data for all flights in
$1^{\circ}$ bins in latitude and normalized by the number of flights still airborne. The
result is shown in Figure~\ref{fig:figure4}, which gives the probability of attaining a
given latitude as a function of time from launch. The solid and dashed lines demarcate 90\%
and 75\% probabilities, respectively, in the following sense: at a given time since launch,
90\% (75\%) of flights would attain latitudes larger (farther south) than the upper solid
(dashed) line, and 90\% (75\%) of flights would attain latitudes smaller (farther north)
than the lower dashed (solid) line. We chose 18 days as the cutoff for these plots, because the number of flights still airborne drops quickly
thereafter. 

An inaccurate interpretation of Figure~\ref{fig:figure4} is to conclude that 80\% of
balloons remained between the two solid contours for the duration of their flights. The
correct interpretation is that 80\% of balloons were between the solid contours at any given
time, though any individual balloon might have wandered in and out of that region repeatedly
during flight. The maximum and minimum latitude a payload may experience up to a certain
point in flight is often of interest and is discussed next. 

\subsection{Probability for southernmost and northernmost latitude} 
\label{sec:min_max_lat}

\begin{figure}
\begin{center}
\includegraphics*[width=10cm]{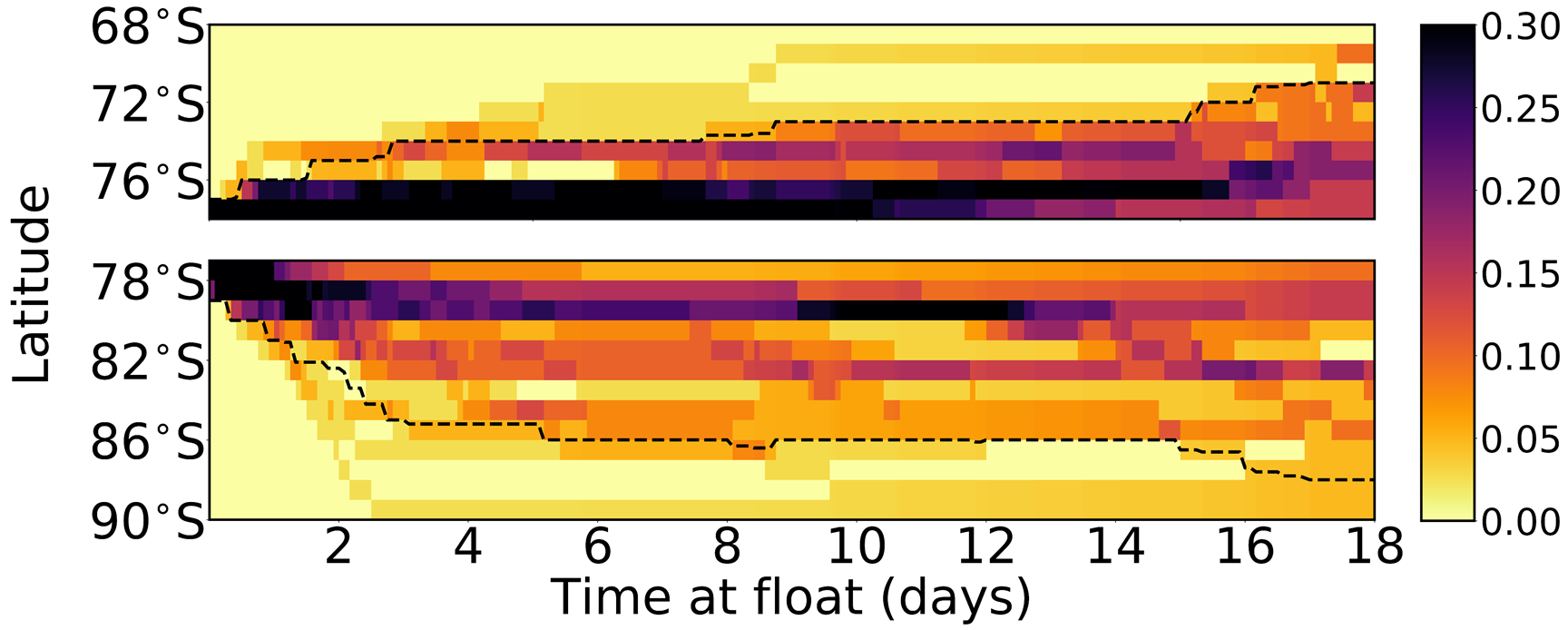}
\end{center}
\caption{Probability for the maximum (upper panel) and minimum (lower panel) latitude a
balloon experienced up to a given time in flight. The dashed contours give the 90th
percentiles in each panel. The color scale saturates at 0.30 to improve contrast; many bins
around the launch location are over-saturated.
\label{fig:figure5}}
\end{figure}

We repeated the analysis described in the previous section, except that for each bin in
time we calculated the northernmost (southernmost) latitude experienced by each balloon
\emph{up to that point in flight}. The result is shown in Figure~\ref{fig:figure5}, again
with 90\% contours shown as dashed black lines. In these plots, for a given length of
flight, 90\% of balloons experienced a northernmost latitude less than the corresponding
value of the contour at that time in the upper panel. Similarly, 90\% of balloons
experienced a southernmost latitude more than the corresponding value of the contour at
that time in the lower panel. This data are also summarized in Table~\ref{tab:table1}.

\subsection{Dependence on Payload Altitude and Launch Date}

Two possible factors that may influence the trajectory of a payload are its float altitude (because wind speed and direction depend on altitude) and the date on which it was launched (because of the seasonal variability of the polar vortex). Given the relatively low number of total flights, we performed a simple cut in each parameter, dividing the flights into two groups about the median altitude and the median launch date, independently. We performed the analysis described in Sections~\ref{sec:lat_vs_time} and \ref{sec:min_max_lat} on each subgroup separately, then performed a Kolmogorov-Smirnov (KS) test to assess the significance of any resulting disparities between the subgroups.

\subsubsection{Latitude dependence on median altitude}

For each flight, we computed a median altitude which we will refer to as the altitude of that flight. The median altitude across all flights was 37.6~km. We will refer to the 20 flights with altitudes below and above this value as the `Low Group' and the `High Group', respectively. These groups had median altitudes of 36.7~km and 38.3~km, respectively. The median flight length within each group was 19 and 16~days, but given the 10~day standard deviation across all flights, this disparity is not significant.

\begin{figure}
\begin{center}
\includegraphics*[width=16cm]{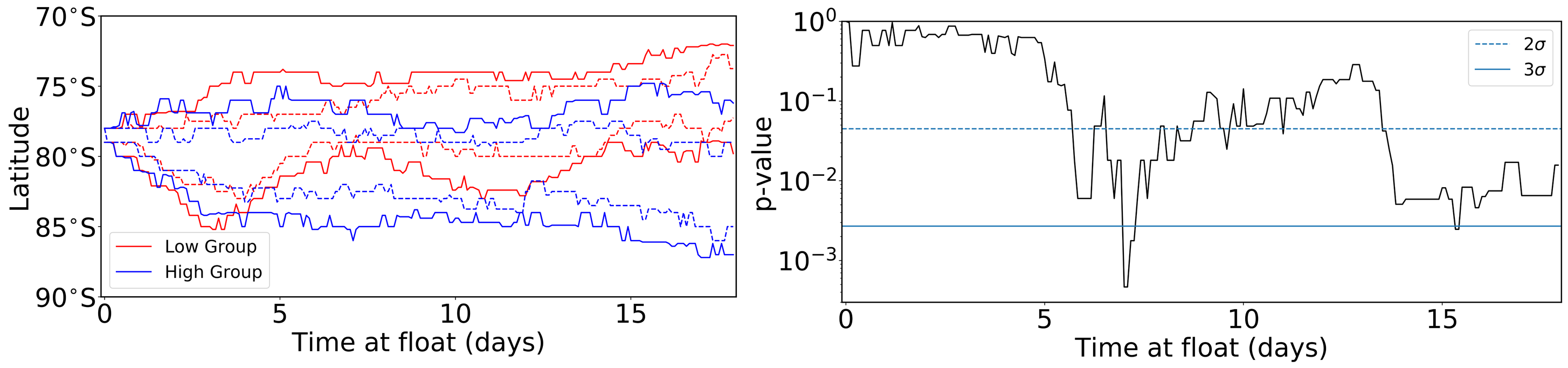}
\end{center}
\caption{(Left) Latitude probability contours as a function of time from launch for flights in the low group (red) and in the high group (blue). The solid contours give the 90th and 10th percentile, and the dashed contours give the 75th and 25th percentile for each temporal bin. (Right) The p-value as a function of time, as determine by the KS test applied to the latitude distributions within each subgroup of flights. This quantifies the probability that the latitude datasets are drawn from a single underlying distribution. 
\label{fig:figure6} }
\end{figure}

We repeated the analysis procedure described in Section~\ref{sec:lat_vs_time} for each subgroup to determine the distribution of latitudes attained by the flights as a function of time. From this, we computed the 10th, 25th, 75th, and 90th percentile contours; the results for each group are shown in the left panel Figure~\ref{fig:figure6}. The flights in the Low Group (red) have a clear tendency to drift north relative to the flights in High Group (blue).

At each temporal bin, we performed the KS test to evaluate the null hypothesis that the two groups were drawn from the same distribution (i.e. that the difference in altitude did not significantly affect the resulting trajectories). A low probability, or small p-value, implies that there is low probability that the groups were drawn from the same underlying distribution; a high p-value neither rejects nor confirms the hypothesis that the two groups were drawn from the same distribution.
This p-value is plotted as a function of time in the right panel of Figure~\ref{fig:figure6}. The p-value is below the 2$\sigma$ threshold for substantial portions and dips below 3$\sigma$ occasionally, suggesting that the disparity between the two groups is caused by the difference in altitudes rather than being a coincidental effect driven by the natural variance in the full dataset. 

\begin{figure}
\begin{center}
\includegraphics*[width=16cm]{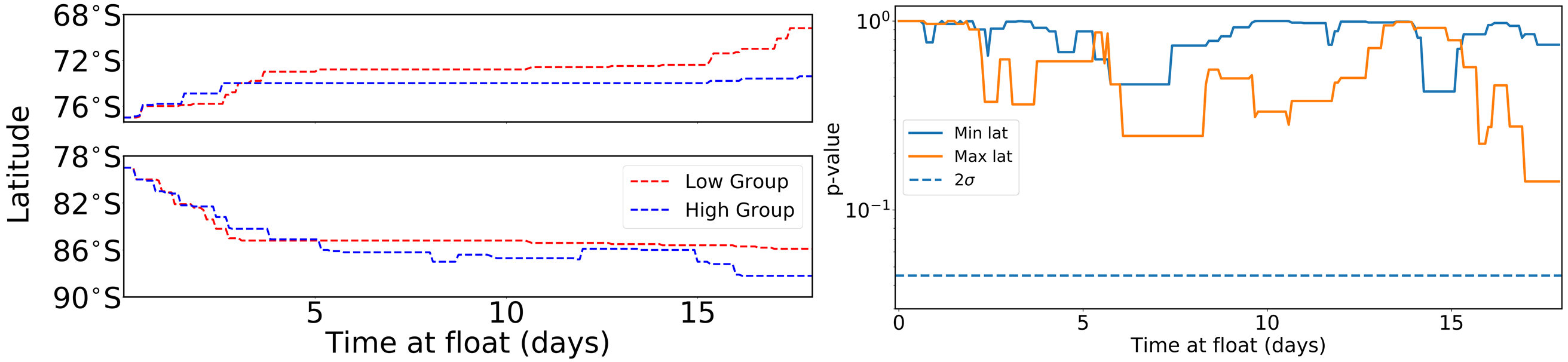}
\end{center}
\caption{(Left) Probability for the maximum (upper panel) and minimum (lower panel) latitude a balloon experienced up to a given time in flight for flights in the low group (red) and in the high group (blue). (Right) Same as Figure~\ref{fig:figure6} for the maximum (orange) and minimum (blue) latitude datasets.
\label{fig:figure7}}
\end{figure}

We repeated the analysis described in Section~\ref{sec:min_max_lat} on each of the two subgroups and found the 90th percentile contours of the minimum and maximum latitude experienced by each flight as a function of time; see the left panel of Figure~\ref{fig:figure7}. These data show a similar trend as Figure~\ref{fig:figure6}, with the higher-altitude flights demonstrating a stronger tendency towards more southerly trajectories. Again, we performed a KS test to assess the significance of the disparity between the two groups; see the right panel of Figure~\ref{fig:figure7}. In this case, we found the p-value never approached the 2$\sigma$ threshold, indicating that any disparity between the two groups was entirely consistent with a single underlying distribution.

\subsubsection{Latitude dependence on launch date}

The median launch date was December 21st;  the 20 flights with earlier (later) launch dates had median launch dates of December~16th and December~29th, and median flight durations of 22 and 16~days, respectively. Applying the KS test to the two sets of durations gave a p-value of 0.06, which is nearly at the $2\sigma = 0.05$ level.

Repeating the latitude dependence analyses from Sections~\ref{sec:lat_vs_time} and \ref{sec:min_max_lat}, we found minimal disparities between the two subgroups; similarly, the KS test applied to the resulting datasets gave no indication that they were drawn from separate distributions. We conclude that while an earlier launch date may be more likely to result in a longer flight, the data do not support any launch date-dependent effects in the latitudes of the resulting trajectories.

\subsection{Uncertainties associated with missing data}

\begin{figure}
\begin{center}
\includegraphics*[width=16cm]{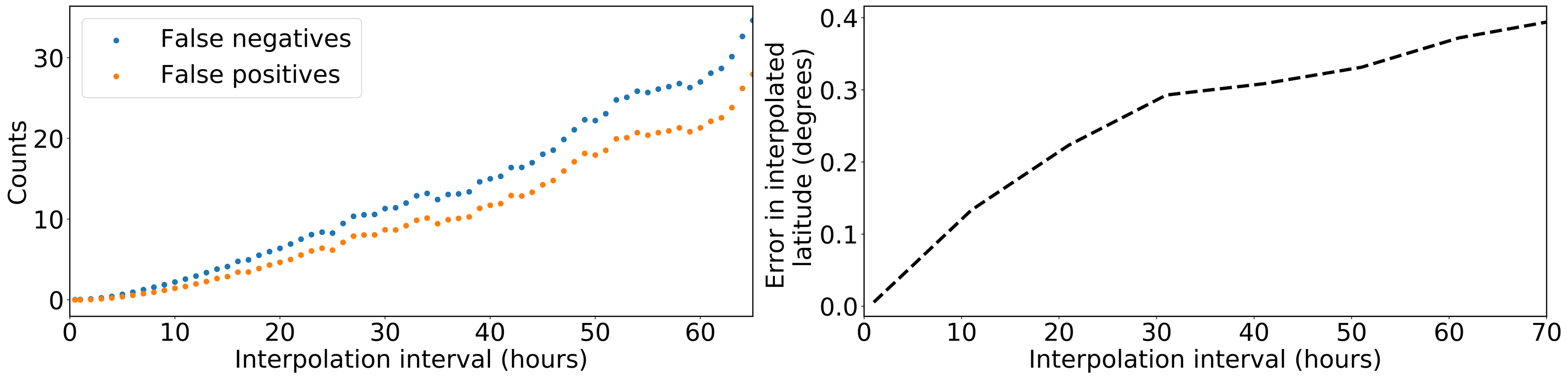}
\end{center}
\caption{(Left) Average number of false negatives (blue dots) and false positives (orange dots) 
associated with linear interpolation over a given time interval. (Right) Expected uncertainty in 
interpolated latitude as a function of length of interpolation interval.
\label{fig:figure8}}
\end{figure}

In the context of the probability hitmap shown in Figure~\ref{fig:figure3}, linear 
interpolation results in both false positives (i.e. bins that appear in the interpolated 
but not in the actual trajectory) and false negatives (i.e. bins that appear in the actual
trajectory but are omitted in the interpolated data). To quantify both these effects, we 
compared interpolated to actual trajectory data for all sections of flights in which no 
data were missing. We determined the expected number of false positives and false 
negatives as a function of interpolation interval, as shown in the left panel of
Figure~\ref{fig:figure8}. Note that the strong correlation between false positives and false negatives is not surprising: when linear interpolation represents a poor approximation of the trajectory, both false positives and false negatives result. Compiling the expected contributions to the total error from all
gaps in the data for which interpolation was required, we expect a total of 270 false positives
and 137 false negatives - as compared to 21,600 total counts. The expected error is therefore
$<$1.5\% with an overall bias of $<$0.7\% towards under-counting the bins that were visited.

For the data presented in Figures~\ref{fig:figure4} and \ref{fig:figure5}, we again 
compared interpolated to actual data for sections of flight with no missing data, this time
in order to assess how the error on the inferred latitude depends on the length of the interpolation
interval. The result, shown in the right panel of Figure~\ref{fig:figure8}, demonstrates that
even for the longest interval for which data is missing (93~hours), the expected error on the
inferred latitude is less than 0.5$^{\circ}$. We therefore expect the resulting error on these data
products due to interpolation over gaps in the data to be negligible.

Five of the flights are also missing data at the end of flight; two of those are also missing the exact 
termination time, though the date is known. These missing data constitute at most 0.3\% of the total
flight time of all 40 flights and were omitted from the analysis presented here.

\section{Discussion and Summary}
\label{sec:conclusion}

For the majority of flights, longitude increased monotonically with time;  therefore, a reasonable alternative would have been to study latitude as a function of longitude. We did not pursue this type of analysis because there is a handful of flights that do not fit the paradigm. For example, they changed direction from west to east, or did not circle the Pole.

\begin{table}
\begin{center}
\caption{Probable northernmost and southernmost latitudes as a function of flight duration.
The values give 90\% confidence intervals. }
\vspace{0.2in}
\begin{tabular}{|c| c c c c c|}
  \hline
  Length of flight (days) & 2 & 8 & 14 & 16 & 18 \\ \hline
  Northernmost latitude (S) & 75$^{\circ}$S & 74$^{\circ}$S & 73$^{\circ}$S & 72$^{\circ}$S & 71$^{\circ}$S \\
  Southernmost latitude (S) & 82$^{\circ}$S & 86$^{\circ}$S & 86$^{\circ}$S & 87$^{\circ}$S & 88$^{\circ}$S \\
  \hline
\end{tabular}
\label{tab:table1}
\end{center}
\end{table}

We analyzed the trajectories of 40 LDB flights from McMurdo station. Some of the data were interpolated, but we expect errors due to interpolation to be less than 2\%. For these 40 flights, we find a median flight duration of 18.7~days. We give the inferred probability of longitude/latitude pixel visitation and the distribution of latitudes as a function of time. We provide the 90th and 75th percentile contours in latitude as a function of flight duration. We give the probable southernmost and northernmost latitude. We find that for the median flight duration of 18~days there is 10\% probability or less that the balloon would drift south of $88^{\circ}$S or north of $71^{\circ}$S; shorter flights are likely to experience smaller ranges in latitude. We investigated the dependence of the trajectories on median altitude; we find that balloons with median altitudes above 37.6~km are less likely to drift northward than balloons with lower median altitudes. 

This is the first time quantitative analysis is carried out using Antarctic balloon flight trajectory data. The results will be useful for scientists planning their missions, and for more informed decision-making during flight. 

\section*{Acknowledgements}
\label{sec:acknowledge}
We thank the Columbia Scientific Ballooning Facility, and particularly Chris Schwantes and
Robert Mullenax, for their assistance in providing and interpreting these data, and we thank the ANITA science team for providing additional trajectory data for their flight. Finally, we thank the anonymous reviewers for their valuable suggestions. 

\section*{Code, Data, and Materials Availability}
 
The raw trajectory data for the 40 flights analyzed here and the data products plotted in Figures~\ref{fig:figure3} through \ref{fig:figure5} are available to the public at https://conservancy.umn.edu/handle/11299/216695.

\bibliographystyle{unsrtnat}
\bibliography{references} 

\end{document}